\documentclass{ws-ijbc}
\usepackage{ws-rotating} 
\begin{document}
       
\catchline{}{}{}{}{} 

\markboth{Zachilas et al}{The phase space of a 4D symplectic map}

\title{The structure of phase space close to fixed points in a 4D symplectic map}
\author{L. ZACHILAS}
\address{Dept. of Economics, University of Thessaly, 43 Korai Str., GR-38333, Volos, Greece
\\zachilas@uth.gr}
\author{M. KATSANIKAS$^{1,2}$ and P.A. PATSIS$^{1}$}
\address{$^{1}$Research Center for Astronomy, Academy of Athens\\
  Soranou Efessiou 4,  GR-11527 Athens, Greece\\
$^{2}$Section of Astrophysics, Astronomy and Mechanics, \\Department of
  Physics, University of Athens, Greece\\
  mkatsan@academyofathens.gr, patsis@academyofathens.gr}

\maketitle

\begin{history}
\received{(to be inserted by publisher)}
\end{history}

\begin{abstract}
We study the dynamics in the neighborhood of fixed points in a 4D symplectic map by means of the color and rotation method. We compare the results with the corresponding cases encountered in galactic type potentials and we find that they are in good agreement. The fact that the 4D phase space close to fixed points is similar to the 4D representations of the surfaces of section close to periodic orbits, indicates an archetypical 4D pattern for each kind of (in)stability, not only in 3D autonomous Hamiltonian systems with galactic type potentials but for a larger class of dynamical systems. This pattern is successfully visualized with the method we use in the paper. 

\end{abstract}

\keywords{Chaos and Dynamical Systems, 4D symplectic maps}

\section{Introduction}
\label{intro}
Patsis and Zachilas [1994] proposed a method to visualize 4D spaces of
section. It is based on rotation of the 3D projections of the figures in order
to understand the geometry  of the  projections  and on color for
understanding the distribution of the consequents in the 4th dimension. This combined information from the rotated, colored, 3D projections allows us to associate specific structures with the dynamical behavior in the neighborhood of periodic orbits.

Recently the method has been applied successfully in a series of papers that studied the dynamics in the neighborhood of periodic orbits in a 3D galactic potential [Katsanikas and Patsis 2011, Katsanikas et al 2011a, Katsanikas et al 2011b]. The results of these papers, together with those of the original one by Patsis and Zachilas [1994], as well as further results from work in progress, find a consistency between the shapes of the encountered structures in the 4D spaces of section and the kind of (in)stability of the periodic orbit. Despite the fact that until now no exceptional behavior has been found, the results cannot be characterized as generic. The up to now studied systems are 3D autonomous Hamiltonians with potentials suitable to describe stellar motion in 3D rotating galactic disks. They can be used to investigate observed morphological features out of the equatorial planes of disk galaxies [Patsis et al 2002]. 

The motivation for investigating the phase space structure of a 4D symplectic map is to check whether or not the behavior encountered in the Hamiltonian systems is model dependent.
So, we apply the same technique in the case of a 4D symplectic map and we study the structure of the phase space in the case of stability and in cases of instabilities studied in the Katsanikas' papers. Our goal is to compare the dynamics of the 4D map with those found in the Hamiltonian system, testing in this way the ubiquity of the results of the latter studies.

The paper is structured as follows: In Sec.~\ref{meth} we briefly describe the method, in Sec.~\ref{map} we present the map we use in our study, in Sec.~\ref{res} we describe our results and finally we enumerate our conclusions in Sec.~\ref{concl}.

\section{A brief description of the method}
\label{meth}
We consider the map $T:\mathbb{R}^4\rightarrow\mathbb{R}^4$ and follow the evolution of the array $(x_1,x_2,x_3,x_4)$.
A set of three coordinates, e.g. $(x_1,x_2,x_3)$, are used
for the 3D projection, while the fourth coordinate (e.g. $x_4$)
determines the color of the consequents. There is a normalization of the color
values in the [min($x_4$), max($x_4$)] interval, which is mapped to
[0,1]. In order to understand the shape of each 3D projection we rotate the figures on the screen of our computer. For presentations on paper we chose a set of projection angles that help the reader understand the shape of the object we describe in the text. We use in our applications the ``Mathematica'' package.
Following the intrinsic ``Mathematica'' subroutines our viewpoint is
given in spherical coordinates. The unit  for the distance $d$ of the
consequents of the surface of section from the observer  is given by
``Mathematica''  in a special scaled  coordinate system, in which the longest
side of the  bounding box has length 1. For all figures  we use  $d=1$. The
method associates  the smooth  distribution  or  the mixing of colors, with
specific types of dynamical  behavior in the 4th  dimension [Patsis and
Zachilas 1994, Katsanikas and Patsis 2011, Katsanikas et al 2011a, Katsanikas et al 2011b].
For a more detailed description see [Katsanikas and Patsis 2011].

\section{The map}
\label{map}

The map we have chosen, $T$, belongs to a family of nonlinear symplectic 4D mappings in $\mathbb{R}^4$ that is a generalization of the standard map. The definition in the general form is:
\begin{eqnarray}
T
\left( 
\begin{matrix} x_1\\x_2\\x_3\\x_4 \end{matrix}
\right) =
\left(
\begin{array}{l}
x_1 + K_1 \sin(x_1 + x_2) + L_1\sin(x_1 + x_2 + x_3 + x_4) \\
x_1 + x_2 \\
x_3 + K_2 \sin(x_3 + x_4) + L_2\sin(x_1 + x_2 + x_3 + x_4) \\
x_3 + x_4 \\
\end{array}
\right) \left(\hspace{-0.4cm}\mod 2\pi\right) 
\end{eqnarray}

Several cases of the maps of this family have been used in the past to study
the dynamics in the neighborhood of fixed points. Already Froeschl\'{e} [1972]
used (1) with $L_1=L_2$ and tried to visualize the 4D surfaces of
section. This was an additional motivation for choosing the particular map to
apply our visualization technique. Furthermore studies accomplished by
Pfenniger [1985], Oll\'{e} and  Pfenniger [1999], Jorba  and Oll\'{e} [2004],
guarantee that Hamiltonian Hopf bifurcations are happening in this system. A 
study of a 4D symplectic map by means of GALI indicators can be found in Manos
et al. [2012]. This allows us to compare also the behavior of the galactic 
type Hamiltonian with that of the map at transitions from stability to complex 
instability. 

Following Pfenniger [1985] and Jorba and Oll\'{e} [2004], we examine the case 
with $L_1+L_2=0$ having as conjugated variables $(x_1,x_2)$ and $(x_4,x_3)$. We also restrict the parameter space by taking $K_2=0$. Then, we rename $L_1=-L_2$ as $L$, and $K_1$ as $K$. It is evident from the stability diagrams in Pfenniger [1985], that there is always a critical value $L_c=-K/4$, for $-8<K<0$, for which we have a transition from stability to complex instability. In order to include the case of such a transition in our calculations, we take here $K=-2$.

We performed our calculations in double, as well as in quadruple precision and we found the same results, except in a case of simple instability (see Sect.~\ref{res} below), where the results are only qualitatively similar. In that case we present the results found with quadruple precision.

\section{4D phase space}
\label{res}

\begin{figure}
\begin{center}
\includegraphics[scale=0.6]{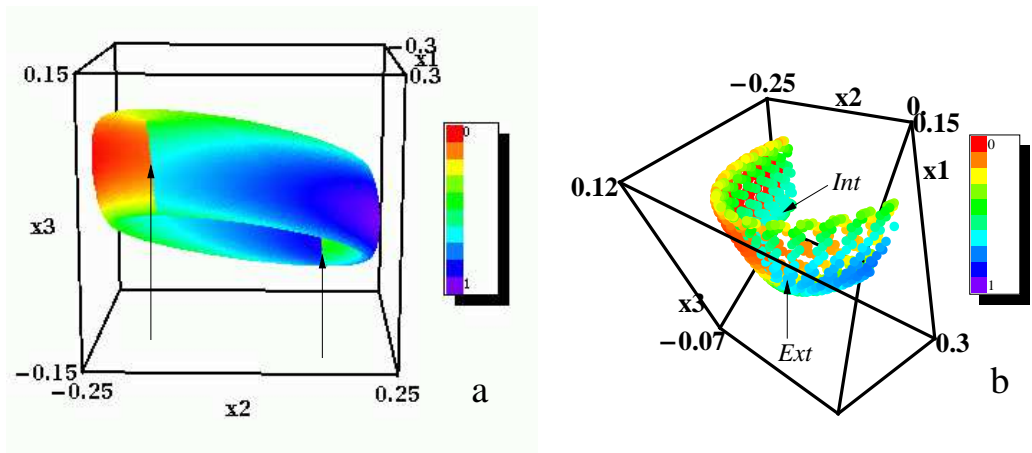}
\caption{The orbital behavior in the neighborhood of a stable fixed point in the $L=0.3$ case. (a) The characteristic toroidal figure associated with stability. The arrows indicate the regions where the smooth color variation changes from the external to the internal side of the torus. The point of view is $(\theta,\phi) = (10^o,270^o)$ (b) An enlargement of (a) at one of the transition regions of the color from the external (``\textit{Ext}'') to the internal (``\textit{Int}'') side of the torus (cf. with figure 12 in Katsanikas and Patsis [2011]). We use less consequents in order to show how the structure of the colors varies. The point of view is now $(\theta,\phi) = (18^o,40^o)$ to show better how the color change side.}
\label{stab}
\end{center}
\end{figure}

We studied the dynamics close to the fixed point (0,0,0,0) by varying $L$.
In the neighborhood of stable fixed points $0<L<0.5$ we find in the 4D phase space invariant tori. Such invariant tori have been presented in projections already by Froeschl\'{e} [1972]. However, with the method of color and rotation we could verify that these tori are 4D objects. Their 3D projections are of the kind called ``rotational tori'' by Vrahatis et al. [1997]. Their 4D representation, in all cases we examined, is in perfect agreement with the rotational invariant tori with smooth color variation on their surfaces found by Katsanikas and Patsis [2011] (see their figure 13). In Fig.~\ref{stab} we give a typical example for $L=0.3$. The initial conditions are $(x_1,x_2,x_3,x_4)=(0.2,0,0,0)$. We use the $(x_1,x_2,x_3)$ 3D projection for the spatial representation of the consequents and $x_4$ to give the color that represents the fourth dimension. In this and in all subsequent figures the color bar on the right hand side gives the range of the color variation that corresponds to the normalized values of the fourth coordinate. Even the change of the smooth color variation from the external to the internal side of the torus described by Katsanikas and Patsis [2011] in their study of the autonomous Hamiltonian system is observed in the case of the rotational torus around the stable fixed point of the symplectic map. The detailed description of the color variation on the surface of the rotational torus should be sought in Katsanikas and Patsis [2011]. These structures characterize the dynamics of ordered behavior in the case of our map. Calculating Lyapunov Characteristic Numbers for all these orbits on the tori we find that they tend very fast to zero.

\begin{figure}
\begin{center}
\includegraphics[scale=0.6]{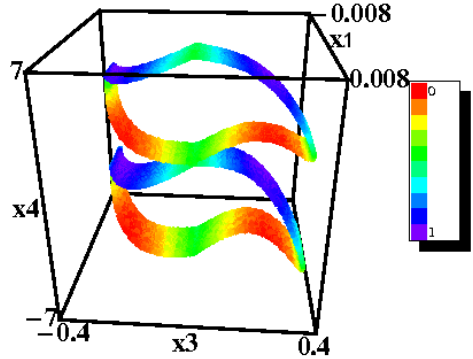}
\caption{The orbital behavior in the neighborhood of a simple unstable fixed point in the $L=-0.03$ case. We observe the characteristic double loop figure associated with simple instability. The point of view is $(\theta,\phi) = (120^o,9^o)$. }
\label{su1}
\end{center}
\end{figure}

Simple instability is encountered for $-4<L<0$  as the $|L|$ value increases. 
In the neighborhood of simple unstable (0,0,0,0) fixed points (two eigenvalues on the unit circle and two on the real axis) we find consequents defining a ribbon of the form of a double loop. This characteristic double loop structure is encountered only close to a critical value of the parameter (the value of $L$) for which the stability of the fixed point changes and it becomes simple unstable.
A typical example for $L=-0.03$ and initial conditions  $(x_1,x_2,x_3,x_4)=(0.001,0,0,-0.001)$ is presented in Fig.~\ref{su1}. It gives  the first $10^5$ consequents. Such structures are associated with simple instability in 3D galactic potentials by Patsis and Zachilas [1994] and are extensively investigated by Katsanikas et al. [2011c].
 The current situation resembles the dynamical behavior encountered in the neighborhood of simple unstable periodic orbits of the z-axis family in 3D rotating Hamiltonian systems [Katsanikas et al. 2011c], when we have a surface of section defined by $z=0$. In the intersection of two branches of the double loop structure we have the same color, indicating that it belongs to the cases where the double loops correspond to a real 4D ``8''-like  structure.
 
\begin{figure}
\begin{center}
\includegraphics [scale=0.65]{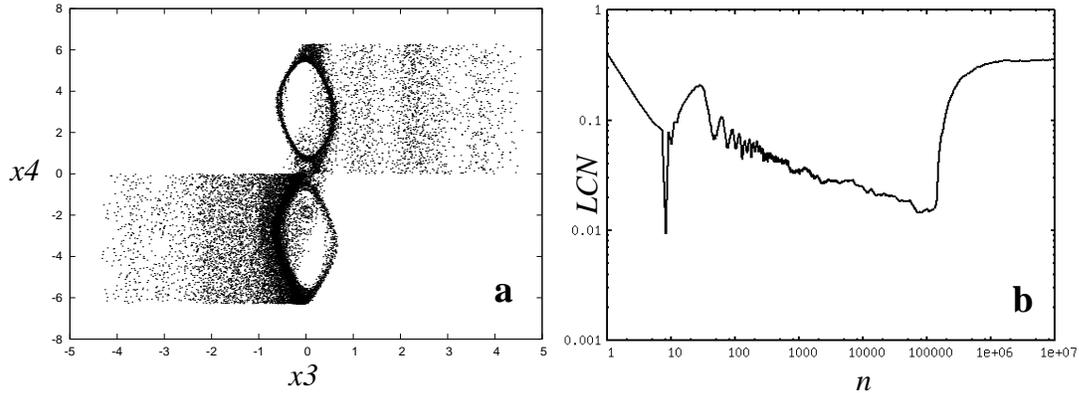}
\caption{(a) The $(x_3,x_4)$ projection of the phase space close to a simple unstable fixed point for $L=-0.1$. The consequents depart from the double loop structure after about $14\times 10^4$ iterations.
(b) 
The $LCN(n)$ evolution. We observe that LCN levels off at a positive value about 0.355.}
\label{u2}
\end{center}
\end{figure}

After a large number of iterations the consequents depart from the surface of the ribbon-like object and build a cloud of points filling finally all available phase space. As $|L|$ increases the consequents leave the ``8''-like  structure faster. In the example of Fig.~\ref{u2}, $L=-0.1$ and the initial conditions are as in the previous case. We observe that the consequents have left the 
``8''-like  structure after $14\times 10^4$ iterations and start visiting larger volumes of the phase space (Fig.~\ref{u2}a).
This is a case where we encountered quantitative differences when we use quadruple precision in our calculations. However, also in this case the results are qualitative the same.
This behavior is depicted in the evolution of the ``finite time'' Lyapunov Characteristic Number $LCN(n) = \left( 1/n\right)  \ln|\xi(n)/\xi(0)|$, where $\xi(0)$ and $\xi(n)$ are the distances of two nearby points at $n=0$ and after $n$ iterations [see e.g. Skokos 2010]. The $LCN(n)$ for the example of Fig.~\ref{u2}a is given in Fig.~\ref{u2}b. The evolution of the $LCN(n)$ index has been followed for $10^6$ iterations. 
The index has an overall decreasing part that ends after about $14 \times 10^4$ iterations (when the consequents start departing from the 4D ``8''-structure). Then it increases and finally it levels off at a value about 0.355. 



\begin{figure}
\begin{center}
\includegraphics [scale=0.6]{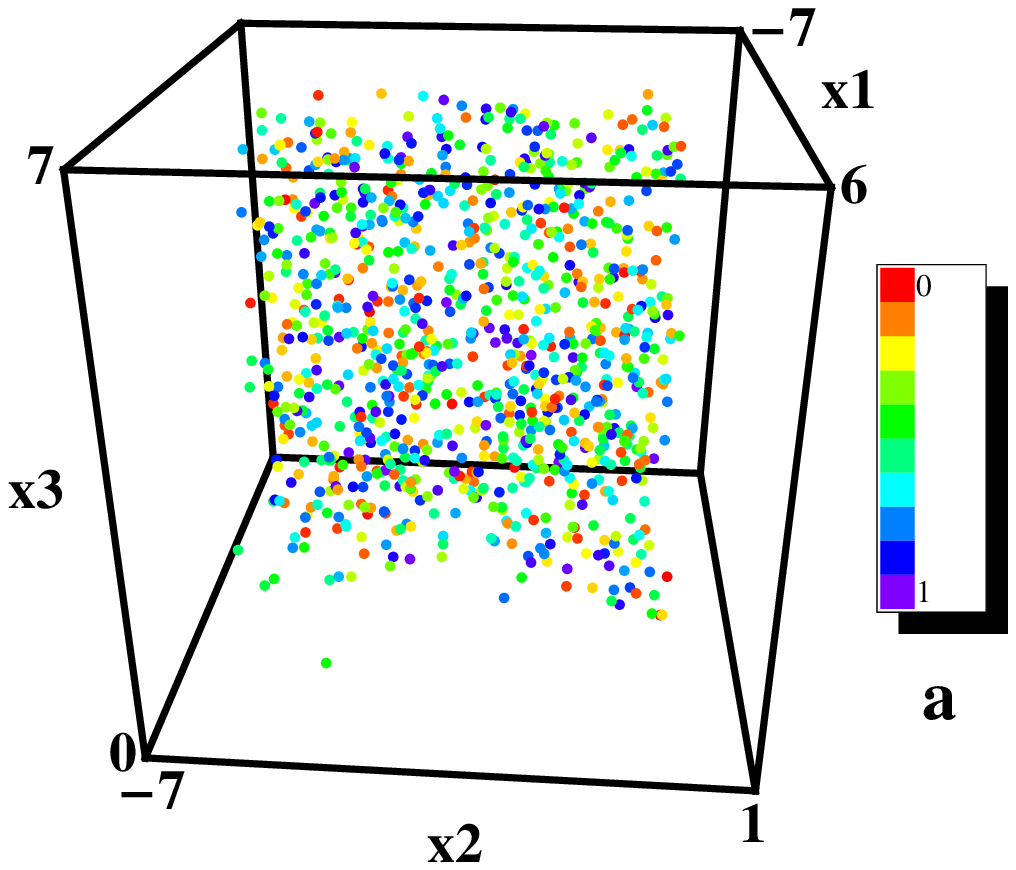}
\includegraphics [scale=0.6]{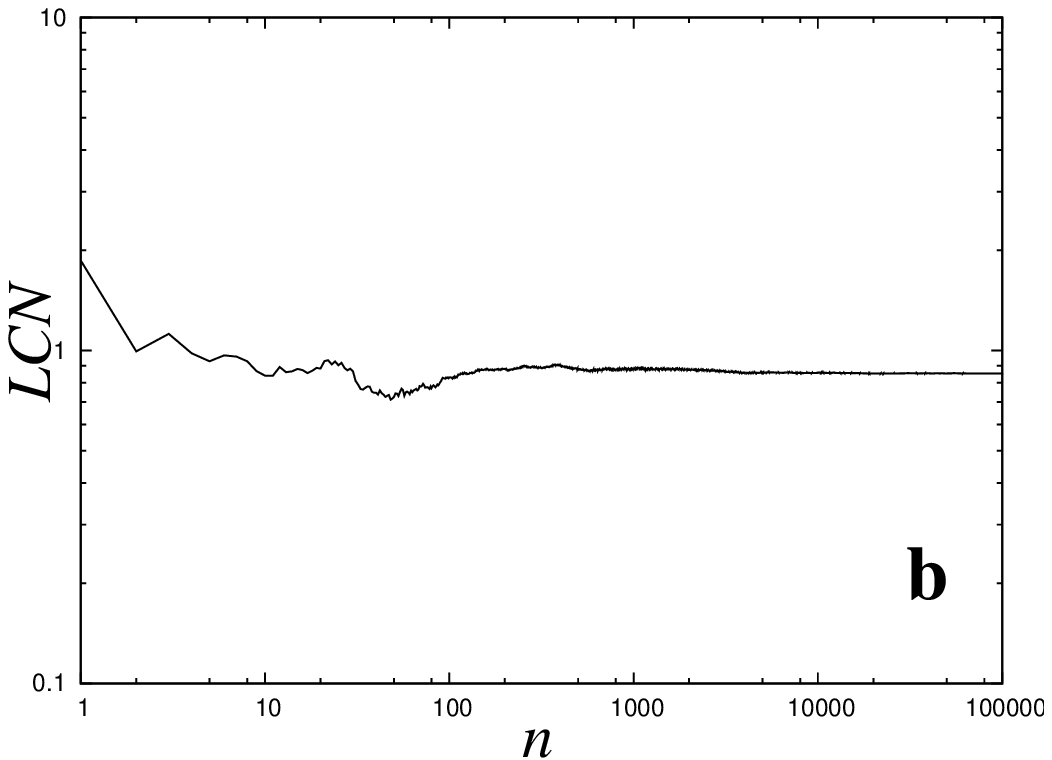}
\caption{a)The orbital behavior in the case of Double instability, for $L=-4.01$. In this case we observe a cloud of points with scattered colors. The point of view is again $(\theta,\phi) = (10^o,120^o)$. (b) The \textit{LCN(n)}}
\label{du}
\end{center}
\end{figure}

The fixed point (0,0,0,0) becomes double unstable for $L<-4$. In all cases we have examined, close to double unstable fixed points we observe clouds of scattered consequents with mixing of colors, even close to the transitions  from simple to double instability. In the example of Fig.~\ref{du}a we give an orbit for $L=-4.01$ with initial conditions $(x_1,x_2,x_3,x_4)=(0.1,0,0,0)$. In this case  
$LCN(n)$ levels off almost immediately at a value about 0.85 (Fig.~\ref{du}b). This is the usual behavior also in most cases studied by Katsanikas et al [2011c] close to double unstable periodic orbits.

\begin{figure}
\begin{center}
\includegraphics[scale=0.5]{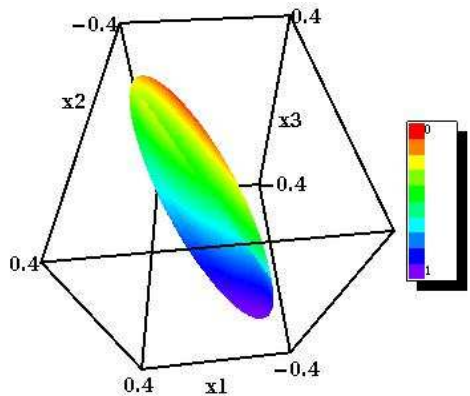}
\caption{The orbital behavior in the neighborhood of a complex unstable fixed point in the $L=0.51$ case. We observe a disky structure with smooth color variation in the 4D phase space of our symplectic map. The point of view is $(\theta,\phi) = (3^o,216^o)$. }
\label{co1}
\end{center}
\end{figure}

\begin{figure}
\begin{center}
\includegraphics [scale=0.6]{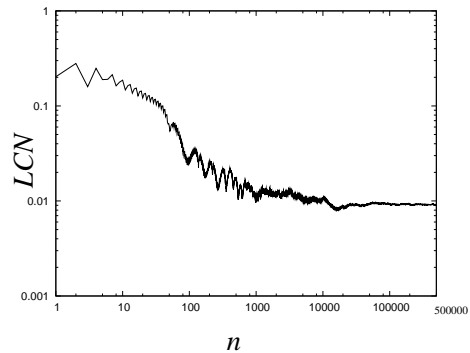}
\caption{The $LCN(n)$ evolution in the case of Fig.~\ref{co1}. We observe that $LCN(n)$ levels off  at a positive value about 0.009, after an initial decreasing part.}
\label{liadisk}
\end{center}
\end{figure}

For  $K=-2$, the eigenvalues of the characteristic polynomial of the monodromy matrix become complex, off the unit circle, for $L>0.5$.  The critical value at which happens the transition from stability to complex instability is $L=0.5$, i.e. for this value we have a Hamiltonian Hopf bifurcation.
Studying the dynamics close to the (0,0,0,0) fixed point, for $L=0.51$, just beyond the critical value, we find the disky structure (confined torus) that characterizes the dynamics close to complex unstable periodic orbits or fixed points. This structure has been found in the 3D projections of the 4D phase space by Pfenniger [1985], Jorba and Oll\'{e} [2004] and Oll\'{e} et al. [2004]. It has been shown by Katsanikas et al. [2011a] that this structure is a 4D object with a smooth color variation as we move from one side of the ``disk'' to the other. For the case of the map (1) we can see it in Fig.~\ref{co1} for $2 \times 10^5$ iterations.
These ``disks'' have an internal spiral structure [Contopoulos et al. 1994, Papadaki et al. 1995, Contopoulos 2002], the 4D structure of which has been shown by Katsanikas et al [2011a]. A similar internal spiral structure as the one described by Katsanikas et al [2011a] in the 4D confined tori (see their figures 2,3,4 and 5) has been encountered in all cases of map (1) we have studied in this paper. In the present case the spiral has six arms.
The behavior of the $LCN$ indices in the two cases are also similar. Fig.~\ref{liadisk} depicts the variation of $LCN(n)$ up to $n=10^5$ and is similar to the $LCN(t)$ variation of the corresponding case of the autonomous Hamiltonian system (cf. figure 6 in Katsanikas et al 2011a).
In both case an initial decrease of the indices is followed by a leveling off.
 
Also the behavior in the neighborhood of a complex unstable fixed point away from the critical $L_c$ value resembles that in the neighborhood of a complex unstable periodic orbit away from the Jacobi constant value for which we have the transition from stability to complex instability. We have clouds of points with mixed colors. In the latter case the $LCN(n)$ index reaches a positive value, larger than that of the confined torus case. The whole variation of the index remains in the cases we examined larger than the values we obtained for the disky structure for the same number of iterations.

\section{Conclusions}
\label{concl}
In this paper we compared the dynamical behavior in the neighborhood of fixed points in a 4D symplectic map and we compared the results with those found in the neighborhood of periodic orbits in a 3D autonomous Hamiltonian system with a potential of galactic type. In our study we used the method of color and rotation. We found that the structures encountered in the spaces of section in the Hamiltonian system, which characterize the kind of (in)stability of the periodic orbits are also found characterizing the dynamics in the neighborhood of fixed points in the 4D symplectic map. These structures are: (1) The tori with the smooth color variation for stability, (2) the double loop ribbons with smooth color variation for the simple unstable case close to transitions from stability to simple instability, (3) the clouds of scattered points in the neighborhood of double unstable periodic orbits,  and (4) the disky structures with smooth color variation close to a complex unstable fixed point near the transition from stability to complex instability. These results indicate that the 4D visualization of the surfaces of section in 3D autonomous Hamiltonian systems and of the phase space in 4D symplectic maps associates the dynamics in the neighborhood of a fixed point, or a periodic orbit, with specific structures.  
We have not found any exception even in the case of the symplectic map we used in the present study. Thus, these structures seem to reflect a generic behavior in a larger class of dynamical systems.
Our study also indicates that the color and rotation method allows us to know the kind of instability of the fixed point by a simple inspection of the 4D phase space and gives in a direct way an insight of the dynamics in such regions, not only in galactic type potentials, but in a broader spectrum of dynamical systems.

\vspace{2cm}
\textit{Acknowledgments}  We thank Prof. Contopoulos for fruitful discussions, as well as both referees for their constructive comments that improved the paper.\\  
 \section{References}
\begin{enumerate}

\item Contopoulos G. [2002] \textit{Order and Chaos in  Dynamical Astronomy}
 Springer-Verlag, New York Berlin Heidelberg.

\item Contopoulos G., Farantos S.C., Papadaki H. and  Polymilis C. [1994]  
 ``Complex unstable periodic orbits and their  manifestation in classical and 
  quantum dynamics'' \textit{Phys. Rev. E} \textbf{50}, 4399-4403.

\item Froeschl\'{e}, C.[1972], ``Numerical Study of a Four-Dimensional Mapping'' \textit{Astron. Astrophys.} \textbf{16}, 172-189

\item Jorba A., Oll\'{e} M. [2004] ``Invariant curves near Hamiltonian Hopf 
  bifurcations of four-dimensional symplectic maps'' \textit{Nonlinearity}  \textbf{17}, 691-710.


\item Katsanikas M., Patsis P.A. [2011] ``The structure of invariant tori in a
  3D galactic potential'' \textit{Int. J. Bif. Chaos} \textbf{21}, No.2,467-496

\item Katsanikas M., Patsis P.A., Contopoulos G. [2011a] ``The structure and evolution of confined tori near a Hamiltonian Hopf Bifurcation'' \textit{Int. J. Bif. Chaos} \textbf{21}, No.8, 2321-2330

\item Katsanikas M., Patsis P.A., Pinotsis A.D. [2011b] ``Chains of rotational tori and filamentary structures close to high multiplicity periodic orbits in a 3D galactic potential''  \textit{Int. J. Bif. Chaos} \textbf{21}, No.8, 2331-2342

\item Katsanikas M., Patsis P.A., Contopoulos G. [2011c] ``Instabilities and Stickiness in a 3D rotating galactic potential'' - submitted.
 \item Manos T., Skokos Ch. and Antonopoulos Ch. [2012] ``Probing the local
   dynamics of periodic orbits by means of generalized alignment index (GALI) 
method'', \textit{Int. J. Bif. Chaos} \textbf{22}, 1250218.
\item Oll\'{e} M., Pfenniger D. [1999] ``Bifurcation at complex instability'' in \textit{Hamiltonian Systems with Three or more Degrees of Freedom}, NATO ASI C, C.Sim\'{o} (ed), pp 518-522, Kluwer, Dordrecht

\item Oll\'{e} M., Pacha J.R. and Villanueva J. [2004] 
  ``Motion close to the Hopf bifurcation of the vertical family of periodic 
  orbits of $L_4$'' \textit{Celest. Mech. Dyn. Astr.} \textbf{90}, 89-109.

\item Papadaki H., Contopoulos G., Polymilis C. [1995] ``Complex Instability'' 
  In: \textit{From Newton to Chaos} ed by A.E. Roy, B.A. Steves, 
 Plenum Press, New York, pg. 485-494.
\item
Patsis P.A., Skokos C., Athanasoula E. [2002] ``Orbital dynamics of three-dimensional bars. - III. Boxy/peanut edge-on profiles'', \textit{Mon. Not. R. Astr. Soc.}, \textbf{337}, 578-596
\item 
  Patsis P.A. and Zachilas L. [1994] ``Using Color and rotation for 
  visualizing  four-dimensional Poincar\'{e} cross-sections:with applications 
  to the orbital  behavior of a three-dimensional Hamiltonian system'' 
  \textit{Int. J. Bif. Chaos}  \textbf{4}, 1399-1424.

\item Pfenniger D. [1985] ``Numerical study of complex instability: I Mappings''  
 \textit{Astron. Astrophys.} \textbf{150}, 97-111. 

\item Skokos Ch. [2010]   ``The Lyapunov Characteristic Exponents and their Computation'', 
\textit{Lect. Not. Phys.} \textbf{790}, 63-135

\item
 Vrahatis, M.N., Isliker, H. \& Bountis, T.C. [1997] ``Structure and breakdown
 of invariant tori in a 4-D mapping model of accelerator dynamics'',
 \textit{Int. J. Bif. Chaos.}  \textbf{7}, 2707-2722.


\end{enumerate}

\end{document}